\documentclass{llncs}
\usepackage{amsmath}
\usepackage{amssymb}
\usepackage{graphicx}
\usepackage{graphics}
\usepackage{algorithm}
\usepackage{algorithmic}
\usepackage{amssymb}
\usepackage{url}
\usepackage{booktabs}
\urldef{\mailsa}\path|{dunan,wubin,wangbai}@bupt.edu.cn|
\urldef{\mailsb}\path|{wangyi.tseg}@gmail.com|
\newcommand{\keywords}[1]{\par\addvspace\baselineskip
\noindent\keywordname\enspace\ignorespaces#1}

\begin{document}

\mainmatter  

\title{Overlapping Community Detection in \\Bipartite Networks\thanks{This work is supported by the National Science Foundation of
China under grant number 60402011, and the National Science and
Technology Support Program of China under Grant No.2006BAH03B05.}}

\titlerunning{Overlapping Community Detection in Bipartite Networks}

%
%
\author{Nan Du\and Bin Wu\and Bai Wang\and Yi Wang\\}
\authorrunning{Nan Du, Bin Wu, Bai Wang, Yi Wang}

\institute{Beijing Key Laboratory of Intelligent Telecommunications Software and Multimedia,\\
Beijing University of Posts and Telecommunications. 100876, Beijing, China\\
\mailsa\\
\mailsb\\
}

%
%

\toctitle{Overlapping Community Detection in Bipartite Networks}
\tocauthor{Nan Du, Bin Wu, Bai Wang, Yi Wang} \maketitle

\begin{abstract}
Recent researches have discovered that rich interactions among
entities in nature and society bring about complex networks with
community structures. Although the investigation of the community
structures has promoted the development of many successful
algorithms, most of them only find separated communities, while for
the vast majority of real-world networks, communities actually
overlap to some extent. Moreover, the vertices of networks can often
belong to different domains as well. Therefore, in this paper, we
propose a novel algorithm \emph{BiTector} (\textbf{Bi}-community
De\textbf{tector}) to efficiently mine overlapping communities in
large-scale sparse bipartite networks. It only depends on the
network topology, and does not require any priori knowledge about
the number or the original partition of the network. We apply the
algorithm to real-world data from different domains, showing that
\emph{BiTector} can successfully identifies the overlapping
community structures of the bipartite networks.
\end{abstract}
\section{Introduction}
In recent years, people have found that both of the physical systems
in nature and the engineered artifacts in human society can be
modeled as complex
networks\cite{watts:smallworld1}\cite{watts:smallworld2}, such as
the internet, the World Wide Web, social networks, citation networks
and etc. Although these systems come from very different domains,
they all have the \emph{\textbf{community}} structure
\cite{newman:community1}\cite{newman:modularity} in common, that is,
they have vertices in a group structure that vertices within the
groups have higher density of edges while vertices among groups have
lower density of edges.

The existence of the community structures has important practical
significance. For example, the communities in World Wide Web
correspond to topics of interest. In social networks, individuals
belong to the same community are probable to have properties in
common. Nowadays, community information is considered to be used for
improving the search engine to provide better personalized results.
Moreover, the information diffusion and spreading mechanism in a
network can be affected and determined by the community structures.
Hence, identifying the communities is a fundamental step not only
for discovering what makes entities come together, but also for
understanding the overall structural and functional properties of
the whole network. As a result, a wide range of successful
algorithms\cite{newman:community2} have been proposed to discover
the community structures. These methods assume that communities are
separated, placing each vertex in only one community. They do not
take into account the possible overlappings\cite{palla:overlapping}
among communities in the real-world scenarios, such as that each of
us may participate in many social cycles according to our various
hobbies.

Moreover, the specific types of the vertices may not belong to the
same domain as well, bringing about a bipartite network structure.
For example, in the scientific collaboration
network\cite{newman2001}, two different types of nodes represent the
authors and papers respectively; in the movie-actor
network\cite{imdb}, each actor is connected to the films where he or
she has starred; in the collaborative recommendation
network\cite{movie}, the edges link each customer to the
corresponding rated or tagged pages, pictures, videos and other
products. In addition, many biological networks are naturally
bipartite, such as the protein interaction network from
yeast\cite{yeast}, where the two types of nodes are bait proteins
and prey proteins, and the human disease network of diseases and
their pathogenicity genes\cite{humandisease}.

Traditionally, the studies of the bipartite networks usually depend
on the one-mode projection of the original network into two
unipartite networks. More specifically, given a bipartite network
$\mathcal {G}_{(U,I)}$, where $U$ and $I$ are the sets of the two
different types of nodes, the one-mode projection converts $\mathcal
{G}_{(U,I)}$ into $\mathcal {G}_{U}$ and $\mathcal {G}_{I}$
respectively. The adjacency matrices of $\mathcal {G}_{U}$ and
$\mathcal {G}_{I}$ are built such that
\begin{displaymath}
E_{ij}(\mathcal {G}_{U}) = \left \{
\begin{array} {l@{\quad}l} 1
& \textrm{if vertex } i \textrm{ and } j \textrm{ have a common
neighbor } i_{k}\in I \textrm{ in } \mathcal
{G}_{(U,I)}\\
0 & \textrm{otherwise}\\
\end{array}\right.
\end{displaymath}
and
\begin{displaymath}
E_{ij}(\mathcal {G}_{I}) = \left \{
\begin{array} {l@{\quad}l} 1
& \textrm{if vertex } i \textrm{ and } j \textrm{ have a common
neighbor } u_{k}\in U \textrm{ in } \mathcal
{G}_{(U,I)}\\
0 & \textrm{otherwise}\\
\end{array}\right.
\end{displaymath}
\begin{figure}
   \centering
   \includegraphics[width=\textwidth, bb=0 0 546 236]{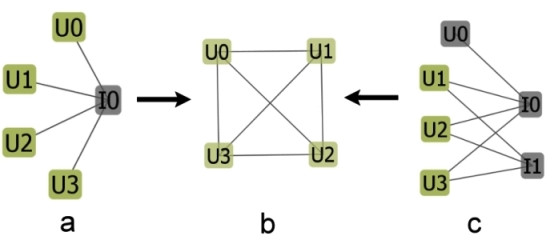}
   \caption{\label{f1}One-mode projection}
\end{figure}

Thus, many existing community detection algorithms can be applied to
$\mathcal {G}_{U}$ and $\mathcal {G}_{I}$ accordingly. Although this
projection approach is simple and intuitive, it may suffer from the
loss of information problem. In general, the real world bipartite
network $\mathcal {G}_{(U,I)}$ is a large sparse graph. However, the
generated graph $\mathcal {G}_{U}$ and $\mathcal {G}_{I}$ may become
very dense as a result of the projection. In Fig.\ref{f1}a, U0, U1,
U2, and U3 have a common neighbor I0, so they form a 4-clique
(complete graph) in Fig.\ref{f1}b by projection. Similarly, we can
also obtain the same 4-clique from Fig.\ref{f1}c. However, it is
easy to see that in Fig.\ref{f1}c, there exists a more closer
relationship among U1, U2, and U3 for that they all have connections
with both I0 and I1. Yet, in Fig.\ref{f1}b, the four nodes are
indistinguishably equivalent with each other. This problem is very
common in real life networks. For example, in collaborative
recommendation network, a very popular film can be rated by hundreds
of users just like the scenario shown in Fig.\ref{f1}a. If we
project the original graph into the network consisting of all the
users, it will contain a huge clique formed by these hundreds of
users. As a result, due to the existence of many superfluous edges
generated by the one-mode projection, the truly meaningful
information may be overwhelmed by the high link density.

Consequently, the main contributions of this paper concentrate on
mining overlapping communities directly on the bipartite networks.
We would like to answer the questions like what groups of people are
interested in what types of products, or what cycles of scientists
prefer to collaborate in what kind of research areas. The rest of
the paper is thus organized as follows: in section 2, we mainly
review some related work. Section 3 describes the overlapping
community detection algorithm \emph{BiTector} in details.
Experimental results and analysis are presented in section 4; and we
conclude the paper in section 5.

\section{Related Work}
One of the classic approaches for detecting community structures in
unipartite networks is the \emph{GN} algorithm\cite{newman:gn} that
introduces a network modularity metric and optimizes it globally to
find the non-overlapping communities.
\textit{Guimer\`{a}}\cite{Guimera:physics0701151} et al. generalizes
this modularity metric to the bipartite networks. They first
differentiate the two parts of the network as the actors and teams,
and then formulate the bipartite modularity from the groups of
actors that are closely interconnected based on joint participation
in many teams. Given vertex $v_{i}$ and $v_{j}$, the bipartite
modularity is defined as the cumulative deviation of the number of
the actual teams where $v_{i}$ and $v_{j}$ have been involved from
the random expectation. Similarly, $Barber$\cite{barber} defines the
bipartite modularity matrix \textbf{B} as an extension of
\emph{Newman}'s recent work\cite{newmanmatrix}. Some key properties
of the eigenspectrum of \textbf{B} are identified and used to
specialize \emph{Newman}'s matrix-based algorithms to bipartite
networks.

In parallel, \emph{Lehmann}\cite{lehmann} et al. extend the
$k$-clique community definition from \emph{Palla}'s
work\cite{palla:overlapping}. They define a $K_{a,b}$ biclique
community as a union of all $K_{a,b}$ bicliques that can be reached
from each other through a series of adjacent $K_{a,b}$ bicliques,
where $a$ and $b$ are the vertices' number belonging to the two
different vertex sets respectively. Just like \emph{Palla}'s work,
two $K_{a,b}$ bicliques are to be adjacent if their overlap is at
least a $K_{a-1,b-1}$ biclique.

To sum up, the modularity-based algorithms, like $GN$ with $O(m^3)$
time complexity ($m$ is the number of edges), are designed to find
the non-overlapping communities and often have the efficiency
problem which makes them unsuitable to the large-scale networks in
practical scenarios. Moreover, the modularity optimization strategy
may introduce a resolution limit\cite{Resolution} as well. For
\emph{Lehmann}'s algorithm, since it extends from \emph{Palla}'s
work, the required user input value $k$, the lower and upper limit
value of the community size, often put a significant impact on the
discovered communities, and are uneasy to be determined before the
algorithm can run. In addition, vertices that are not included in
any $K_{a,b}$ bicliques will be ignored, so the set of all the
detected communities usually can not cover all the vertices of the
original graph.

Therefore, to overcome these shortages, we propose \emph{BiTector}
by a local optimization strategy, which does not suffer from the
resolution problem, and does not require any priori knowledge about
the community's number or other related thresholds to assess the
community structure. As of this writing, \emph{BiTector} is the
first method that can handle bipartite networks consisting of
millions of nodes and edges.

\section{BiClique-based Overlapping Community Detection Algorithm}
Instead of dividing a network into its most loosely connected parts,
\emph{BiTector} identifies the communities based on the most densely
connected parts, namely, the \textbf{\emph{bicliques}}. We treat
each group of highly overlapping maximal bicliques as the clustering
cores. Surrounding each core, we build up the communities in an
gradually expanding way according to certain metrics until each
vertex in the network belongs to at least one community.
\subsection{Notations and Definitions}
In this paper, we consider simple and connected graphs only, i.e.,
the graphs without self-loops or multi-edges. Given graph $\mathcal
{G}_{(U,I)}$, where $U$ and $I$ or $U_{\mathcal {G}}$ and
$V_{\mathcal {G}}$ are the sets of the two different types of nodes,
$V(\mathcal {G}_{(U,I)})=U_{\mathcal {G}}\cup I_{\mathcal {G}}$ and
$E(\mathcal {G}_{(U,I)})$ denote the sets of all its vertices and
edges respectively.
\begin{definition}
Given sub-bigraph $S_{(U,I)}, U_{S}\subseteq U_{\mathcal{G}},
I_{S}\subseteq I_{\mathcal{G}}$, if $\forall u_{i}\in U_{S},
v_{j}\in I_{S}$, $\exists e_{(u_{i},v_{j})}\in E(S_{(U,I)})$, then
$S_{(U,I)}$ is a biclique. If there is no any other biclique
$S'_{(U,I)}$, such that $U_{S}\subset U_{S'}$ and $I_{S}\subset
I_{S'}$, $S_{(U,I)}$ is called the maximal bicliques.
\end{definition}
\begin{definition}For a given vertex $v$, $N(v)=\{u|(v,u)\in
E(G)\}$, we call $N(v)$ is the neighbor set of $v$. For sub-bigraph
$S_{(U,I)}$, $N(U_{S})=\bigcup N(u_{i})-I_{S},u_{i}\in U_{S}$, and
$N(I_{S})=\bigcup N(v_{j})-U_{S}, v_{j}\in I_{S}$, $N(U_{S})\cup
N(I_{S})$ is called the neighbor set of $S_{(U,I)}$.
\end{definition}
\begin{definition}
$\mathcal{M}(\mathcal {G}_{(U,I)})$ denotes the set of all maximal
bicliques $B_{(U,I)}$ $(|U_{B}|\geq 2, |I_{B}|\geq 2)$ in $\mathcal
{G}_{(U,I)}$. Given vertex $v_{i}\in V(\mathcal {G}_{(U,I)})$,
$\mathcal{B}(v_{i})\subseteq \mathcal{M}(\mathcal {G}_{(U,I)})$ is
the set of all maximal bicliques that contain $v_{i}$. The set of
all $\mathcal{B}(v_{i})$ is denoted as $\mathcal{B}$. For any pair
of sub-bigraph $\mathcal{G}_{i}$ and $\mathcal{G}_{j}$, a
\emph{Closeness Function} $isClose(\mathcal{G}_{i},\mathcal{G}_{j})$
is defined and implemented in the next section to identify whether
they could be merged together by quantifying how "close" they
actually can be. Given any two maximal bicliques $B_{m},B_{n}\in
\mathcal{B}(v_{i}), |U_{B_{m}}|>|U_{B_{n}}|$, $\mathcal{G}_{m}$ and
$\mathcal{G}_{n}$ are the bi-subgraphs induced on $B_{m}$ and
$B_{n}$ respectively. If $isClose(\mathcal{G}_{m},\mathcal{G}_{n})$
returns \emph{true}, we say $B_{n}$ is contained by $B_{m}$, denoted
by $B_{n}<B_{m}$. If $B_{m}$ is not contained by any other maximal
bicliques in $\mathcal{B}(v_{i})$, $B_{m}$ is called the
$\mathbf{core}$ and the set of all cores is denoted by $\mathcal
{C}$.
\end{definition}
\begin{definition}
Let $S_{0}$,$S_{1}$,...,$S_{n-1}$ be the sub-bigraph of $\mathcal
{G}_{(U,I)}$ such that $V(S_{0})\cup$,...,\\$V(S_{n-1})=V(\mathcal
{G}_{(U,I)})$. For any pair of $S_{i}$ and $S_{j}$, if
$|E(S_{i})|>|E_{between}(S_{i},S_{j})|$, $S_{i}$ is defined as the
\emph{\textbf{community}} of $\mathcal {G}_{(U,I)}$.
\end{definition}
\subsection{Algorithm}
\textit{BiTector} first enumerates all maximal bicliques in
$\mathcal {G}_{(U,I)}$. Because a maximal biclique is a complete
sub-bigraph, it is thus the densest community structure which can
represent the closest relationship between the two types of vertices
in the given network. Given two sub-bigraphs $\mathcal{G}_{i}$ and
$\mathcal{G}_{j}$, the basic idea of the \emph{closeness function}
$isClose(\mathcal{G}_{i},\mathcal{G}_{j})$ depends on the link
pattern between $\mathcal{G}_{i}$ and $\mathcal{G}_{j}$ to quantify
the influence that they put on each other. We use $\Delta_{ij}$ to
denote the common vertices between $U_{\mathcal{G}_{i}}$ and
$U_{\mathcal{G}_{j}}$, and $\Gamma_{ij}$ to denote the common
vertices between $I_{\mathcal{G}_{i}}$ and $I_{\mathcal{G}_{j}}$
accordingly. The left sub-bigraph of $\mathcal{G}_{i}$ is then
defined as $\mathcal{L}_{i}$ with
$U_{\mathcal{L}_{i}}=U_{\mathcal{G}_{i}}-\Delta_{ij}$ and
$I_{\mathcal{L}_{i}}=I_{\mathcal{G}_{i}}-\Gamma_{ij}$. Similarly,
the left sub-bigraph of $\mathcal{G}_{j}$ is denoted as
$\mathcal{L}_{j}$ with
$U_{\mathcal{L}_{j}}=U_{\mathcal{G}_{j}}-\Delta_{ij}$ and
$I_{\mathcal{L}_{j}}=I_{\mathcal{G}_{j}}-\Gamma_{ij}$. We define the
sub-bigraphs induced on $U_{\mathcal{L}_{i}}\cup
I_{\mathcal{L}_{j}}$, and $U_{\mathcal{L}_{j}}\cup
I_{\mathcal{L}_{i}}$ are
$\mathcal{G}_{(U_{\mathcal{L}_{i}},I_{\mathcal{L}_{j}})}$ and
$\mathcal{G}_{(U_{\mathcal{L}_{j}},I_{\mathcal{L}_{i}})}$
accordingly. Here the influence that $\mathcal{G}_{i}$ puts on
$\mathcal{G}_{j}$ is defined based on $U_{\mathcal{G}_{i}}$. It is
equivalent if we start from $I_{\mathcal{G}_{i}}$.
\begin{displaymath}
inf_{ij} =
|E(\mathcal{G}_{(U_{\mathcal{L}_{i}},I_{\mathcal{L}_{j}})})|+|E(\mathcal{G}_{(U_{\mathcal{L}_{j}},I_{\mathcal{L}_{i}})})|
- |E(\mathcal{L}_{j})|
\end{displaymath}
It is apparent that for $\mathcal{G}_{i}$ and $\mathcal{G}_{j}$,
$inf_{ij}$ actually reflects the number of edges between them minus
that of $\mathcal{G}_{j}$'s inner edges. If both $inf_{ij}\geq 0$
and $inf_{ji}\geq 0$, then $\mathcal{G}_{i}$ and $\mathcal{G}_{j}$
should be merged together as a single graph; otherwise, they will be
separated apart. The implementation of
$isClose(\mathcal{G}_{i},\mathcal{G}_{j})$ is formulated in
Algorithm \ref{a1}.

\begin{figure}
   \centering
   \includegraphics[width=0.8\textwidth, bb=0 0 508 186]{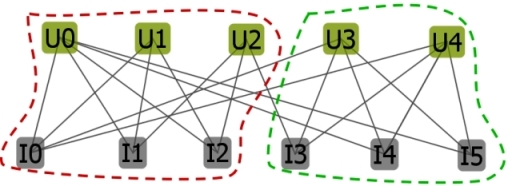}
   \caption{\label{f2}Example of Core Formation}
\end{figure}

To make things more concrete, an illustrated example is given on the
network shown in Fig.\ref{f2}. There are two sub-bigraphs:
$\mathcal{G}_{0}$ cycled by red dashed-line, and $\mathcal{G}_{1}$
cycled by green dashed-line. $U_{\mathcal{G}_{0}}=\{U_{0}, U_{1},
U_{2}\}$. $I_{\mathcal{G}_{0}}=\{I_{0}, I_{1}, I_{2}\}$.
$U_{\mathcal{G}_{1}}=\{U_{3}, U_{4}\}$.
$I_{\mathcal{G}_{1}}=\{I_{3}, I_{4}, I_{5}\}$. $inf_{01} = 2 + 2 - 6
= -2 < 0$, and $inf_{10} = 2 + 2 - 8 = -4 < 0$, so $\mathcal{G}_{0}$
and $\mathcal{G}_{1}$ should not be merged together.
Starting from $\mathcal{M}(\mathcal {G}_{(U,I)})$, we first find
$\mathcal{B}(u_{i})\in \mathcal{B}$ for every vertex $u_{i}\in
\mathcal {G}_{(U,I)}$. Because every maximal biclique in
$\mathcal{B}(u_{i})$ corresponds to one group of vertices in
$U_{\mathcal{G}}$ which together with $u_{i}$ are closely
interconnected based on the jointly connections with certain cluster
of vertices in $I_{\mathcal{G}}$, $\mathcal{B}(u_{i})$ covers all
the densest communities where $u_{i}$ has participated. $\forall
u_{i},u_{j}\in U_{\mathcal{G}}$, $\mathcal{G}_{i}$ and
$\mathcal{G}_{j}$ represent the sub-bigraphs induced on
$\mathcal{B}(u_{i})$ and $\mathcal{B}(u_{j})$. If
$isClose(\mathcal{G}_{i},\mathcal{G}_{j})$ returns true, which means
all or most of $u_{j}$'s relationships are covered by those of
$u_{i}$, $u_{j}$ should thus stay in the same community with
$u_{i}$. We rearrange the elements of $\mathcal{B}$ according to the
descending order of $|\mathcal{B}(u_{i})|$. Let
$|\mathcal{B}(u_{k})|$ be the element of $\mathcal{B}$ whose size is
the largest. We put $|\mathcal{B}(u_{k})|$ to set $\mathcal {H}$ and
removed it from $\mathcal{B}$. All the other elements contained by
$|\mathcal{B}(u_{k})|$ are also removed from $\mathcal{B}$. Again,
we pick the next largest element of $\mathcal{B}$, put it to
$\mathcal{H}$, removed it as well as those elements it contains from
$\mathcal{B}$. The process is continued until $\mathcal{B}$ is
empty, so set $\mathcal{H}$ stores the elements being independent of
each other.
\begin{algorithm}
\caption{\label{a1}$isClose(\mathcal{G}_{i},\mathcal{G}_{j})$}
\begin{algorithmic}[1]
\STATE\COMMENT{ $\mathcal{G}_{L} \Leftarrow \mathcal{G}_{i}$ or
$\mathcal{G}_{j}$ with the larger set of $U$, $\mathcal{G}_{S}
\Leftarrow \mathcal{G}_{i}$ or $\mathcal{G}_{j}$ with the smaller
set of $U$}
\STATE\COMMENT{$\Delta =U_{\mathcal{G}_{L}}\cap
U_{\mathcal{G}_{S}}$, $\Gamma =I_{\mathcal{G}_{L}}\cap
I_{\mathcal{G}_{S}}$}
\IF{$U(\mathcal{G}_{S})\subseteq
U(\mathcal{G}_{L})$ or $I(\mathcal{G}_{S})\subseteq
I(\mathcal{G}_{L})$) or $(I(\mathcal{G}_{L})\subseteq
I(\mathcal{G}_{S}))$}
\STATE return true
\ELSE \STATE
$C_{L}=|E(\mathcal{G}_{(\Delta,\Gamma)})| - |E(\mathcal{L}_{L})|$
\STATE $C_{S}=|E(\mathcal{G}_{(\Delta,\Gamma)})| -
|E(\mathcal{L}_{S})|$ \IF{$(C_{L}\geq 0)$ or $(C_{S}\geq 0)$ or
($inf_{LS}\cdot inf_{SL}\geq 0)$}
\STATE return true
\ENDIF \ENDIF
\STATE return false
\end{algorithmic}
\end{algorithm}
\begin{figure}
   \centering
   \includegraphics[width=0.8\textwidth, bb=0 0 515 162]{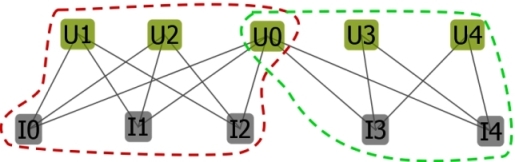}
   \caption{\label{f3}Example of Core Formation}
\end{figure}

In general, the distribution of the vertex degree in bipartite
networks conforms to a power-law. It is common that a few vertices
in $U_{\mathcal{G}}$ have connections with nearly all vertices in
$I_{\mathcal{G}}$. As a result, these vertices can appear in lots of
maximal bicliques repeatedly. For example, in Fig.\ref{f3}, we can
see that
$\mathcal{B}(U0)=\{\{U0,U1,U2,I0,I1,I2\},\{U0,U3,U4,I3,I4\}\}$, and
$U0$ has connections with all the vertices from $I0$ to $I4$.
However, it is obvious that $\{U0,U1,U2,I0,I1,I2\}$ and
$\{U0,U3,U4,I3,I4\}$ are two different communities with a
overlapping vertex $U0$. Consequently, to address this problem,
given $u_{i}$ and $\mathcal{B}(u_{i})$, we need to further refine
$\mathcal{B}(u_{i})$ into several sub-bigraphs representing
different communities that $u_{i}$ has taken part in simultaneously.
Therefore, for each element $H_{k}\in \mathcal{H}$, every maximal
biclique $B_{m}\in H_{k}$ is sorted by the descending order of
$U_{B_{m}}$. Given $B_{m},B_{n}$ with $|U_{B_{m}}|\geq|U_{B_{n}}|$,
if $isClose(\mathcal{G}_{m},\mathcal{G}_{n})$ returns true, $B_{n}$
is thus contained by $B_{m}$. If $B_{m}$ is not contained by any
other elements in $H_{k}$, $B_{m}$ is regarded as the \textbf{core},
and will be put in set $\mathcal{C}$. This process is continued
until every element in $\mathcal{H}$ has been refined. The whole
procedure is described in algorithm \ref{a2}.

\begin{algorithm}
\caption{\label{a2}$CoreFormation(\mathcal{M}(\mathcal
{G}_{(U,I)}))$}
\begin{algorithmic}[1]
\STATE $\mathcal{C} \Leftarrow \emptyset$, $\mathcal{H} \Leftarrow
\emptyset$, $contained \Leftarrow \emptyset$ \STATE Get
$\mathcal{B}$ from $\mathcal{M}(\mathcal {G}_{(U,I)})$ and sort
$\mathcal{B}(u_{i})\in \mathcal{B}$ by the descending order of
$|\mathcal{B}(u_{i})|$ \FOR{$\forall \mathcal{B}(u_{i})\in
\mathcal{B}$} \IF{$\mathcal{B}(u_{i}) \not\in contained$} \STATE
$contained \Leftarrow \mathcal{B}(u_{j})$, if
$isClose(\mathcal{G}_{i},\mathcal{G}_{j})$ returns true \STATE add
$\mathcal{B}(u_{i})$ to $\mathcal{H}$ \ENDIF \ENDFOR \FOR{$\forall
H_{k}\in \mathcal{H}$} \STATE $contained \Leftarrow \emptyset$
\FOR{$\forall B_{m}\in H_{k}$} \IF{$B_{m} \not\in contained$} \STATE
$contained \Leftarrow B_{n}$, $B_{n}<B_{m}$ \STATE add $B_{m}$ to
$\mathcal{C}$ \ENDIF \ENDFOR \ENDFOR \STATE return $\mathcal{C}$
\end{algorithmic}
\end{algorithm}
\subsubsection{Clustering}
Once all the cores have been detected, we carry out a clustering
process to associate the left vertices to their "closest" cores. For
each sub-bigraph $\mathcal{G}_{i}$ induced on $C_{i}\in \mathcal
{C}$, we gradually expand $\mathcal{G}_{i}$ by adding the vertices
in set $N(U_{\mathcal{G}_{i}})\cup N(I_{\mathcal{G}_{i}})$. Given
vertex $\forall v_{i}\in V(\mathcal {G}_{(U,I)})$ and $\forall
C_{j}\in \mathcal{C}$, the distance between $v_{i}$ and
$\mathcal{G}_{j}$ is defined as follows:
\begin{displaymath}
\mathcal{D}(v_{i},\mathcal{G}_{j})= \left \{
\begin{aligned}
         \frac{|N(v_{i})\cap I_{\mathcal{G}_{j}}|}{|N(v_{i})\cup I_{\mathcal{G}_{j}}|} &, v_{i}\in U_{\mathcal{G}} \\
         \frac{|N(v_{i})\cap U_{\mathcal{G}_{j}}|}{|N(v_{i})\cup U_{\mathcal{G}_{j}}|} &, v_{i}\in I_{\mathcal{G}} \\
\end{aligned}
\right.
\end{displaymath}
As a consequence, $v_{i}$ is assigned to the cores with the maximum
distance value. Since that any vertex might have the same maximum
distance value with more than one core, $v_{i}$ can thus be assigned
to multiple cores simultaneously.

Since that for vertex $v_{i}$, it actually does not have connections
with all the cores in $\mathcal{C}$. Therefore, we adopt a coloring
strategy to reduce the computation cost. First, the vertices covered
by all cores in $\mathcal{C}$ are colored as \emph{old}. We use set
$U_{\mathcal{C}}\subseteq U_{\mathcal{G}}$ and
$I_{\mathcal{C}}\subseteq I_{\mathcal{G}}$ to store the two types of
vertices covered by $\mathcal{C}$. Next, every new vertex in
$N(U_{\mathcal{C}})$ and $N(I_{\mathcal{C}})$ is assigned to its
closest cores, and colored as \emph{old}. As a result, every core is
now expanded. Again, starting from $N(U_{\mathcal{C}})$ and
$N(I_{\mathcal{C}})$, all new vertices that have not been colored in
$N(N(U_{\mathcal{C}}))$ and $N(N(I_{\mathcal{C}}))$ are going to be
assigned and colored. The clustering continues until all the
vertices of the network are colored as \emph{old}.
\begin{algorithm}
\caption{\label{a3}$Clustering(\mathcal{C})$}
\begin{algorithmic}[1]
\FOR{$C_{i}\in \mathcal{C}$} \STATE $\forall v_{k}\in
V(\mathcal{G}_{C_{i}})$ is marked as \emph{old} \ENDFOR \STATE
$U_{Exp}\Leftarrow N(I_{\mathcal{C}})$, $I_{Exp}\Leftarrow
N(U_{\mathcal{C}})$ \WHILE{\emph{not all vertices in
$V(\mathcal{G}_{(U,I)})$ are colored}} \FOR{$\forall v_{i}\in
U_{Exp}\cup I_{Exp}$} \IF{$v_{i}$ is not colored} \STATE assign
$v_{i}$ to its closest core, and color $v_{i}$ as \emph{old} \ENDIF
\ENDFOR \STATE $U'_{Exp} \Leftarrow \emptyset$, $I'_{Exp} \Leftarrow
\emptyset$ \STATE add $v_{j}$ to $U'_{Exp}$, if $\forall v_{j}\in
N(I_{Exp})$ and $v_{j}$ is not colored \STATE add $v_{k}$ to
$I'_{Exp}$, if $\forall v_{k}\in N(U_{Exp})$ and $v_{k}$ is not
colored \STATE $U_{Exp} \Leftarrow U'_{Exp}$, $I_{Exp} \Leftarrow
I'_{Exp}$ \ENDWHILE \STATE sort $Core$ according to descending order
of $|U_{\mathcal{G}_{C_{i}}}|, C_{i}\in Core$ \FOR{$C_{i}\in
\mathcal{C}$} \IF {$C_{i}$ is not merged} \STATE $C_{j}$ is merged
to $C_{i}$, if $isClose(\mathcal{G}_{i},\mathcal{G}_{j})$ returns
$true$. \ENDIF \ENDFOR
\end{algorithmic}
\end{algorithm}
In the end, let $\mathcal{C}'$ denote the set of every expanded
core. We use the same process as the \emph{Core Formation} to
compare the closeness between $C'_{i}$ and $C'_{j}\in \mathcal{C}'$.
If $isClose(\mathcal{G}'_{i},\mathcal{G}'_{j})$ returns $true$,
$C'_{i}$ and $C'_{j}$ is merged together. The whole process is
presented in algorithm \ref{a3}.
\subsection{Complexity}
Like the classic maximal clique problem in unipartite network, the
enumeration of all maximal bicliques is a NP problem as well.
However, for most real world bipartite networks, they are often
large sparse graphs$(N=|V(\mathcal{G})|, M=|E(\mathcal{G})|,
N\approx M)$, and there exist modern algorithms that are very
efficient on sparse graphs. Because the enumeration of maximal
bicliques is equivalent to the \emph{Closed Item Set} problem, we
use the \emph{LCM} (Linear time Closed itemset Miner)\cite{LCM} to
mine all the maximal bicliques. On sparse graphs, the computational
complexity of \emph{LCM} is almost proportional to $O(M)$. The
calculation of set $\mathcal{H}$ costs $O(N)$. Let $S_{C}$ be the
maximum size of $\mathcal{B}(u_{i})\in \mathcal{B}$. It costs
$O(|\mathcal{H}|\times S_{C})$ to calculate the core set
$\mathcal{C}$. In the end, the clustering process costs $O(N\times
|\mathcal{C}|)$. Because on sparse bipartite networks
$|\mathcal{M}(\mathcal {G}_{(U,I)})|\approx N \approx M$,
$|\mathcal{H}|\ll N$, $S_{C}<|\mathcal{C}|\ll|\mathcal{M}(\mathcal
{G}_{(U,I)})|$, the total complexity of $BiTector$ is therefore
$O(M^{2})$.
\section{Experimental Results}
In this section, we will present the experimental results and
analysis on several real, large bipartite networks from different
domains. All experiments are done on a single PC (3.0GHz processor
with 2Gbytes of main memory on Linux AS3 OS). The execution time of
\emph{BiTector} includes both of the \emph{\textbf{biclique finding
time}} and \textbf{\emph{the community detection time}}. The
experimental results are shown in Table \ref{t1}.
\begin{table}
\caption{\label{t1}Experimental Results} \centering
\begin{tabular}{l c c c}
\toprule
Graph & Vertices $|V|$ & Edges $|E|$ & Time(s)\\
\midrule \texttt{DAVIS SOUTHERN CLUB WOMEN}\cite{WOMEN} & 32 & 93 & 0.5\\
\texttt{NATION-SPORT NETWORK OF OLYMPIC GAMES}\cite{olympic} & 515 & 208 & 1\\
\texttt{CUSTOMER-PRODUCT NETWORK}\cite{kdd2000} & 2008 & 3258 & 2\\
\texttt{PROTEIN INTERACTION NETWORK OF YEAST}\cite{CCNR} & 3,740 & 4,480 & 2 \\
\texttt{AUTHOR-PAPER NETWORK OF arXiv}\cite{arxiv} & 20,454 & 24,154 & 6\\
\texttt{MOVIE-RATING NETWORK OF NETFLIX}\cite{netflix} & 75,179 & 100,000 & 92\\
\texttt{BOOK-RATING NETWORK}\cite{book} & 263,804 & 433,695 & 4,028\\
\texttt{IMDB NETWORK}\cite{CCNR} & 289,435 & 637,035 & 4,312\\
\bottomrule
\end{tabular}
\end{table}
\subsubsection{\texttt{DAVIS SOUTHERN CLUB WOMEN.}} 
\begin{figure}
   \centering
   \includegraphics[width=0.8\textwidth, bb=0 0 437 262]{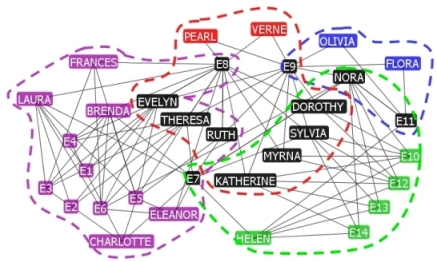}
   \caption{\label{f4}SOUTHERN CLUB WOMEN}
\end{figure}

The Southern women data set describes the participation
of 18 women in 14 social events. The women and social events
constitute a bipartite network; an edge exists between a woman and a
social event if the woman was in attendance at the event. This data
set have been much studied by \emph{Davis} as part of an extensive
study of class and race in the Deep South. $BiTector$ finds 4
overlapping communities shown in Fig.\ref{f4}.
Each community is circled by one colored dashed-line, and the
overlapping vertices are colored by \emph{Black}. It is apparent
that $E8$ and $E9$ are two very famous clubs attracting 9 women to
join. Similarly, \emph{Barber}'s algorithm also gets 4 separated
communities, while \textit{Guimer\`{a}}'s method finds two coarse
ones.
\subsubsection{\texttt{CUSTOMER-PRODUCT NETWORK}} is
derived from the purchase data of \emph{Gazelle.com}, a legwear and
legcare web retailer that closed their online store on 8/18/2000. An
edge connects a customer to the products he or she has ordered.
\subsubsection{\texttt{PROTEIN INTERACTION NETWORK OF YEAST}}
contains two types of proteins. One represents the bait proteins and
the other represents the prey proteins. An edge links a prey protein
to a bait protein if the prey protein binds to the bait one.
\subsubsection{\texttt{AUTHOR-PAPER NETWORK OF arXiv}} 
\begin{figure}
   \centering
   \includegraphics[width=\textwidth, bb=0 0 524 180]{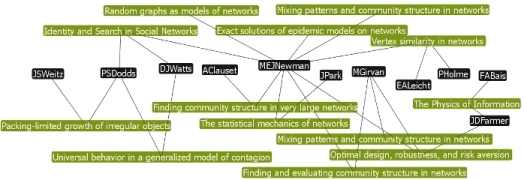}
   \caption{\label{f5}Experts with Research Areas}
\end{figure}

presents the
relationships among authors and papers. An edge links an author to a
paper if this author has published the paper before. Each discovered
community in this network can intuitively links certain experts to
their research areas that are reflected by the published papers on
which they have once collaborated. Fig.\ref{f5} describes one
community where Prof. \emph{M.E.J. Newman} has been involved.
\emph{Newman} has proposed the classic $GN$
algorithm\cite{newman:gn} for community detection in unipartite
networks, and the community detected by $BiTector$ in Fig.\ref{f5}
can directly finds one of the circles where he has been often
involved in the physics society.
\subsubsection{\texttt{MOVIE-RATING NETWORK OF NETFLIX}} is composed
of users and their rated movies. \emph{Netflix} provides an
evaluation mechanism that enables users to rate movies from score 0
to score 10 to express their preferences. There exists an edge
between a user and a movie if this user has rated the movie. In our
experiment, we build the network from \emph{Netflix}'s rating data
in 2006.
\subsubsection{\texttt{BOOK-RATING NETWORK}} is built from the \emph{Book-Crossing}
community. In our experiment, there exists en edge between a user
and a book if this user has given a non-zero rating score to the
book.
\subsubsection{\texttt{IMDB NETWORK}} is composed of actors and
movies. A link connects an actor or actress to a movie he or she has
once starred.
\subsubsection{}In the experiments, except for the
\texttt{DAVIS SOUTHERN CLUB WOMEN}, both \emph{Barber}'s and
\textit{Guimer\`{a}}'s algorithms are not suitable to run on the
other datasets within the acceptable time. For \emph{Lehmann}'s
algorithm, since the discovered communities depend on the user input
value $k$, and the required lower bound and upper bound of the
community size, we do not include the correspondent results here. We
further evaluate the homogeneity of $BiTector$'s discovered
communities by comparing them with their counterparts in the random
bipartite networks.
\begin{figure}
  \centering
   \includegraphics[width=0.8\textwidth, bb=0 0 328 279]{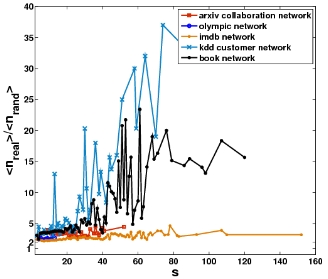}
   \caption{\label{f6} Communities' Homogeneity}
\end{figure}
For any discovered community $C_{(U,I)}$, we first randomly choose
$|U_{C}|$ vertices from $U_{\mathcal{G}}$ into set $U_{R}$. Then
from the union neighbor set of the chosen vertices, we further
randomly choose $|I_{C}|$ vertices into set $I_{R}$. As a
consequence, we obtain a randomly generated community $R_{(U,I)}$
having the same size with $C_{(U,I)}$. In Fig.\ref{f6}, each symbol
corresponds to the average number of the inner edges for a given
community size, $n_{<real>}$, divided by the same quantity found in
random sets, $n_{<rand>}$. We can see that the
$n_{<real>}/n_{<rand>}$ ratio is significantly larger than 1,
indicating that the communities discovered by \emph{BiTector} tend
to contain closely interrelated entities, a homogeneity that
supports the validity and effectiveness of the discovered
communities.
\subsubsection{\texttt{NATION-SPORT NETWORK OF OLYMPIC GAMES}}.
Besides the bipartite networks we just discussed above, $BiTector$
is further challenged on the networks of Olympic Games in Summer
from 1896 to 2004. In each year, we build the network according the
relationships between nations and the correspondent sports. An edge
links a nation and a specific sport if the nation has won medals in
that sport. There are totally 25 networks being built from 1896 to
2004. The average number of nodes and edges are 515 and 208
respectively.

Each discovered community directly represents certain group of
sports in which a few nations often compete with each other.
\begin{figure}
   \centering
   \includegraphics[width=0.9\textwidth, bb=0 0 535 397]{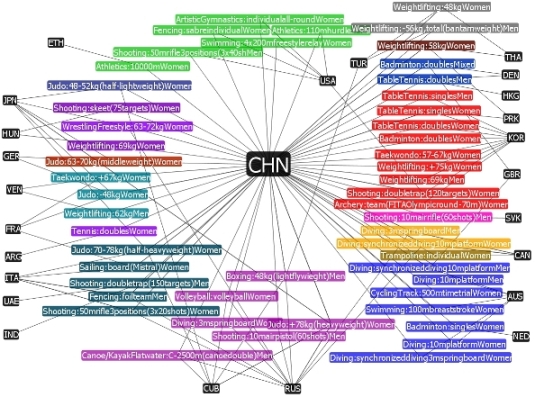}
   \caption{\label{f7}China in Olympic Games 2004}
\end{figure}
\begin{figure}
   \centering
   \includegraphics[width=\textwidth, bb=0 0 594 406]{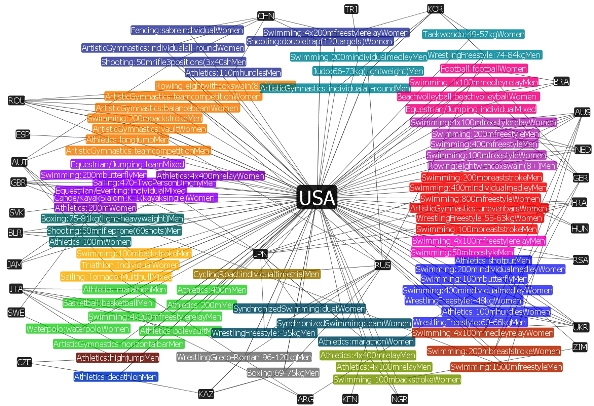}
   \caption{\label{f8}USA in Olympic Games 2004}
\end{figure}
For example, Fig.\ref{f7} depicts the sports in which \emph{China}
has won medals as well as the correspondent competitive nations in
the year 2004. Each community that \emph{China} has been involved
are marked as different colors. It is very intuitive that in the
sports such as \emph{TableTennis}, and \emph{Badminton}, \emph{KOR}
is a strong competitor, while in \emph{Swimming} and \emph{Diving},
\emph{China} has to compete with \emph{USA} and \emph{AUS}.

By contrast, Fig.\ref{f8} presents the sports in which \emph{USA}
has won medals in the year 2004. It is apparent that most of
\emph{USA}'s advantage sports concentrate on \emph{swimming},
\emph{Athletics}, as well as \emph{Gymnastics} with its major
competitors such as \emph{AUS} in \emph{swimming}, \emph{ROU} in
\emph{Gymnastics}, and \emph{ITA} in \emph{Athletics}.
\begin{figure}
   \centering
   \includegraphics[width=\textwidth, bb=0 0 520 436]{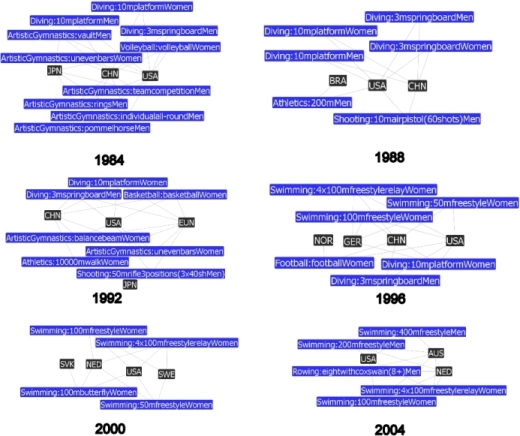}
   \caption{\label{f9}CHN vs. USA in Olympic Games From 1984 to 2004}
\end{figure}

Given the set of communities at time $t$, $\mathcal {C}_{t}$, for
any community $\mathcal {C}_{t}^{i}\in \mathcal {C}_{t}$, if there
exists at least one community $\mathcal {C}_{t+1}^{j}\in \mathcal
{C}_{t+1}$, such that
\begin{displaymath}
\frac{|E(\mathcal{G}_{\mathcal {C}_{t}^{i}})\cap
E(\mathcal{G}_{\mathcal {C}_{t+1}^{j}})|}{|E(\mathcal{G}_{\mathcal
{C}_{t}^{i}})\cup E(\mathcal{G}_{\mathcal {C}_{t+1}^{j}})|}\geq f
\end{displaymath} we say $\mathcal {C}_{t+1}^{j}\in \mathcal
{C}_{t+1}$ is the $descender$ of $\mathcal {C}_{t}^{i}\in \mathcal
{C}_{t}$, and $\mathcal {C}_{t}^{i}\in \mathcal {C}_{t}$ evolves to
$\mathcal {C}_{t+1}^{j}\in \mathcal {C}_{t+1}$. In our experiments,
the empirical value of $f$ on the olympic data is set to 0.1.
Fig.\ref{f9} depicts the evolving trace of one community where
\emph{CHN} competes with \emph{USA} in the sports of \emph{Diving}
and \emph{ArtisticGymnastics} from 1984 to 2004. We see that
although \emph{CHN} has competed with \emph{USA} in
\emph{Diving:10mplatformWomen} continuously for 4 Olympic Games,
\emph{USA} has still been keeping its advantage in \emph{water
sports} steadily.
\section{Conclusion}
In this paper, we have proposed a new method \textit{BiTector} for
efficient overlapping community identification in large-scale
bipartitie networks. We have demonstrated the effectiveness and
efficiency of \textit{BiTector} over a number of real networks
coming from disparate domains whose structures are otherwise
difficult to understand. Experimental results show that this
algorithm can extract meaningful communities that are agreed with
both of the objective facts and our intuitions. \textit{BiTector}
avoids loss of essential information caused by the one-mode
projection approach and the thresholding procedures, and is expected
to be of great help in many practical scenarios.
\subsubsection*{Acknowledgments.}
We thank \emph{Xin Yang} for the collection of the Olympic Games
data greatly.

\end{document}